# Hiding in the Clouds and Building a Stealth Communication Network

Wei Jiang*    Adam Bowers†    Dan Lin‡

Department of Electrical Engineering and Computer Science
University of Missouri
Columbia, MO 65211, USA

*wjiang@missouri.edu
†acbqbd@mail.missouri.edu
‡lindan@missouri.edu

**Abstract**

Social networks, instant messages and file sharing systems are common communication means among friends, families, coworkers, etc. Due to concerns of personal privacy, identify thefts, data misuse, freedom of speech and government surveillance, online social or communication networks have provided various options for a user to guard or control his or her personal data. However, for most these services, user data are still accessible by the service providers, which can lead to both liability issues if data breach occurs at the server side and data misuse by the network administrators. To prevent service providers from accessing user data, secure end-to-end user communication is a must, like the one provided by WhatsApp. On the other hand, the services provided by such communication network can still be interfered by an authority. For a communication network to be stealthy, the following features are essential: (1) oblivious service, (2) complete user control and flexibility, and (3) lightweight. In this paper, we first discuss the features and benefits of a stealth communication network (SNET), and then we propose a theoretical framework that can be adopted to implement an SNET. By utilizing the framework and the existing publicly available cloud storages, we present the implementation details of an instance of SNET, named Secret-Share. Last but not least, we discuss the current limitations of Secret-Share and its potential extensions.

## 1   Introduction

Communication is an essential part of our daily lives. In today's digital world, there are many online tools or platforms that allow people to conveniently contact their families, friends and coworkers to share about their personal lives or work related activities. Although tools like social networks, emails and cloud storage facilitate our communication need, they also open the doors for exposing a person's private life, stealing personal identifiable information, and being watched by the Big Brother. Since a social network, by definition, collects personal information, much of which is contextually private, this has created a burgeoning privacy problem.

One may argue that since existing social network providers are well-established IT companies, they can be trusted. However, this is a questionable assumption at best. For example, Facebook keeps some user data indefinitely even if the account is deleted, tracks user movements across the web, used 'Likes' in ads, and gave data to the government [17]. Indeed, many media organizations, including the Wall Street Journal, have been arguing that Facebook is probably selling user profile data to third-party advertisers to make huge profits [1, 10, 19]. Furthermore, as the recent Facebook and Cambridge Analytica [15] shows, the *users* of the social network have very little control over what happens to them. Privacy is also at risk due to the third-party applications running on the social network that keep tracking user network data.

Facebook and other social networks have addressed the privacy problem by giving users more and more control over their information and how it is used. Nevertheless, since user data are stored on the network server, the data are still accessible by the network administrator, which could lead to liability issues if data breach occurs at the server and data misuse by the administrator. In addition, a third party authority may ban the use of social media to control public opinions which has very negative impacts on the freedom of speech. Although online social networks are generally owned by private companies, an authority can force them to release private user data and expose political views of certain group of users. For instance, Lavabit was shut down because it refused to provide the authority the private key used to secure user emails [23]. As a result, it is essential to have a stealth, both oblivious and secure, communication network that will not be easily controlled by an authority. We term such network as an SNET.

According to the previous discussions, it is apparent that to prevent service providers and other entities from accessing user data and stopping secure communication services, the following features are necessary for an SNET:

1. *User controlled data encryption*: User data stored and managed by the service providers should be encrypted with the user's private key, and only the user can decrypt the data.

2. *Secure end-to-end communication*: The communication channel between any two users, e.g., Alice and Bob, should be secured. That is, only Alice and Bob can access the information shared between them.

3. *Oblivious communication*: To prevent an authority from forcing the service providers to disclose user information, the operations of private communications between any two users should not be known or easily detectable. Ideally, the service providers should not even know if they have actually provided secure communication services to a group of users.

In addition to privacy and availability, the following two properties are essential from usability perspective:

4. *Lightweight computation*: Most computations are done by the network provider, and the computation on the client side can be performed by a device with limited storage, memory and computing power.

5. *Transparent and non-intrusive*: The privacy-preservation mechanism of SNET does not interference with user common activities within a communication network, and it is not noticeable by any users.

There have been efforts to create secure communication platforms, such as Persona [4] and Diaspora [9], among others. Persona utilizes attribute based encryption (ABE [14]) to control access to a user's data. The user needs to perform very expensive ABE operations; thus, the system is not suitable for mobile device with limited storage and computing power. Diaspora is a distributed implementation of an online social network, and it mainly prevents a single service provider from taking full control of the user data. The data hosted by individual servers are not encrypted which still opens the door for data misuse and security breaches. Detailed comparisons of the existing secure communication platforms with our proposed SNET are given in Section 2. In summary, none of the existing solutions satisfy any of the following properties: efficient end-to-end communication, service obliviousness, or lightweight computation on the client side.

## 1.1 Technical Limitations of the Existing Solutions

To achieve certain degree of privacy-protection, the system design among the existing systems utilizes either one or both of the following strategies:

- Distributed architecture: Under this approach, user data are distributed among multiple servers. For example, a user may have multiple pictures, and each picture could be stored on one of the servers. It is also possible that a single picture is broken into multiple pieces, each of which is stored at a server. The design is very similar to BitTorrent [5]. This approach does not allow a single server to have a complete control over the user data.

- End-to-end encryption: This approach offers the best security guarantee by establishing a secure communication channel between any two users. As a result, the service providers have no way to access user data.

The two strategies have some major limitations to prevent the existing systems from becoming an SNET:

- Distributed architecture: Each data piece leaks some or all the information regarding the original data. Without adopting end-to-end encryption, this approach does not provide sufficient security or privacy.

- End-to-end encryption: The computation cost on the client/user side can be very high for sharing information among a large group of users.

More importantly, the existing solutions cannot efficiently and effectively achieve the two key guarantees for an SNET, oblivious and secure end-to-end communication, simultaneously. Next we preview our technical contributions in developing an SNET, especially achieving efficient oblivious and secure end-to-end communication.

## 1.2 Our Main Contributions

To achieve oblivious communication, it is apparent that a distributed architecture is needed in order to make tracking or identifying the service providers very hard. However, a fully distributed peer-to-peer network has several disadvantages:

- Increased complexity of secure end-to-end communication.

- Prone to failures at each peer, and trusting every peer to provide reliable services may be unrealistic.

- The peer knows it is part of an SNET.

To improve the degree of obliviousness and reduce the chance of intervention by a third party authority, it is necessary that even the service providers do not actually know they are performing the SNET services. This seems impossible to obtain; nevertheless, the following observation can lead to a feasible solution:

- **Key observation to achieve obliviousness** - If an SNET can be seamlessly integrated into the existing and commonly used communication services, it becomes difficult for the service providers to detect the existence of the SNET.

Furthermore, to prevent an authority to shut down all communication services in hope of eliminating an SNET, we need to utilize services that are prevalent in our daily activities. For instance, it is unlikely to shut down all social media, cloud storage, and public forums without affecting the general public.

Therefore, to achieve the best service performance, availability and obliviousness, a novel system architecture is needed. In this paper, our main contribution is to propose a system architecture that can be used to develop a stealth (oblivious and secure) communication platform SNET that possesses the aforementioned features. Specific contributions include:

- A semi-distributed system architecture that enables secure data management and efficient, secure, end-to-end user communications. This architecture can be seamlessly integrated into the existing cloud storage and social media services in order to make the communication oblivious.

- A system prototype of SNET, named Secret-Share, under the proposed system architecture to demonstrate its feasibility and practicality in the real world.

- Analyzing alternative designs that trade-off between obliviousness and ease of use, and other potential security issues and mitigation approaches to prevent servers to identify the existence of Secret-Share.

## 1.3 Organization

The rest of the paper is organized as follows: Section 2 presents the existing secure communication platforms and discusses their limitations, Section 3 details the proposed semi-distributed system architecture for implementing an SNET, the threat models and the security guarantees of the SNET, Section 4 presents system prototypes of SNET, termed as Secret-Share and Secret-Share* that trade-off between ease of use and obliviousness according to the threat models, Section 5 discusses the security and performance of Secret-Share and Secret-Share*, Section 6 presents potential attacks to our current implementations and suggests possible counter measures to eliminate or mitigate these attacks, and Section 7 concludes the paper with future research directions.

# 2 Related Work

There are many apps out there to connect users and allow them to chat. However, in exchange, these products tend to keep personal data about their users that they might use in questionable ways. Take for instance, the use of Facebook data by Cambridge Analytica to help advertisers target citizens in the 2016 election [22]. In this attack a significant number of individuals didn't even consent for 3rd parties to receive their personal data. There have been plenty of leaks or abuse of individuals data like this on other social media platforms [2, 18].

## 2.1 Private Messaging Platforms

Given these attacks, there is a need for social media and messaging platforms that either do not store user data or cannot access them if they do store the data. One alternative is Diaspora [9]. Files are encrypted via AES [21], and use standard public key exchange protocol to share the encryption keys. Instead of a centralized server, each user selects several pods to store their information on. Pods are nodes on the distributed network hosted by someone using the service. These keep users from having to trust a centralized server, and pods are not able to see encrypted files. However, the user still has to trust the pod to function according to the protocol. Additionally, encryption for each user with whom a message is shared adds a higher overhead for sharing and messaging. Another alternative is SecuShare [26]. It is similar in that it encrypts messages so only intended recipients can see them. In addition, it uses GNUnet [12] so that network traffic is hidden. This prevents outsiders from monitoring traffic to induce who is talking to whom; however, SecuShare is not oblivious to the servers running it.

As opposed to those approaches, RetroShare [25] has more features similar to a typical social media platform. Instead of encrypting messages to share with contact individually, a user builds a trusted network of friends. Each friend belongs to various user-defined groups and can only access messages and posts for their group. While this approach uses encryption in a similar fashion, it is a peer-to-peer network and has issues concerning security and practicality since users need a large network of friends they trust to help distribute their files. In addition, there is the popular WhatsApp [34] which has the option to encrypt all communications between any two users with AES. However, it uses one central server to help connect people and exchange keys to make messaging possible. The issue with one centralized server is even if it does not know the private key to encrypt contents stored on the server, it still knows who is communicating with whom and does not offer oblivious communication.

A full list of these secure communication systems can be found in [35, 27], and they vary from the degrees of privacy protection to usability-driven functionalities. In summary, the existing implementations to achieve secure communication generally combine distributed architectures and end-to-end encryptions. Peer-to-peer architecture allows fully distributed management of user data. Based on this architecture, an intermediate peer who is responsible of storing and sending other users' data may know the data if end-to-end secure communication is not implemented. In conclusion, the existing secure communication systems lack the following two necessary properties of an SNET: (1) efficient secure end-to-end communication and (2) being oblivious.

## 2.2 Tor and Other Secure File Sharing Tools

The main goal of Tor [31] is to allow users to anonymously connect to remote servers by preventing network traffic analysis. Tor can also help users to visit blocked websites and contents. Tor alone will not lead to a secure and oblivious communication network, but it can be adopted in our proposed architecture to protect user anonymity and prevent linking attacks. Detailed discussion is provided in Section 5.

However, Tor itself has vulnerabilities to traffic analysis and other attacks as described in [20]. There are works that attempt to overcome some of these problems with respect to messaging, specifically [16, 32, 33]. They use private key encryption to send messages between users, much like the Private Messaging Platforms described previously. The novel contribution of these works is to hide metadata that an adversary could learn if messages were sent by Tor or other approaches. However, even with this increased security it still needs a server to facilitate communication between parties. Furthermore, there is even more overhead in this approach due to shuffling and noise used to hide metadata. Therefore, it has too high of overhead for practical use and the servers cannot be made oblivious to the communication.

There are tools to secure remote cloud based file storages, such as Boxcryptor [7] and Sookasa [30]. They allow users to encrypt their data stored on a cloud, but they are not designed for message sharing. More importantly, these tools still lack efficient ways to provide secure end-to-end communication and to make their services oblivious.

Table 1: Common Notations

| Symbol | Definition |
| --- | --- |
| SNET | Stealth (secure and oblivious) communication network |
| TPA | Third party authority |
| SP | Service provider |
| Secret-Share | An implementation of SNET |

## 3 System Architecture of SNET

As discussed in Section 1 and Section 1.1, none of the existing secure communication networks simultaneously satisfy: (1) user controlled data encryption, (2) oblivious and efficient secure end-to-end communication, and (3) lightweight computation, and the existing system architectures listed in [27, 35] cannot be adopted to efficiently and effectively implement these features. As a result, we propose a novel semi-distributed system architecture which takes advantage of both fully distributed and centralized system architectures. A set of commonly used notations in this paper are given in Table 1.

### 3.1 An Overview of SNET

The intuition behind the proposed system design is as follows. For the application to be lightweight, most computations need to be done by a central server. If there is only one server, then symmetric key encryptions, such as AES [21], must be used to establish end-to-end secure communications between any two users, but this approach does not work efficiently (concrete illustrations are given later). To enable both centralized computing and end-to-end secure communication, the proposed system architecture utilizes a semi-distributed design:

- The architecture adopts two or more servers. Most computations occur at the servers, and the application running on the user side is lightweight.

- The architecture utilizes a secret sharing scheme [13] to establish secure end-to-end communications, offering at least the same security as AES based implementations.

The design of the architecture is inspired by the concept of Secure Multiparty Computation (SMC) [13, 36] and can be used to implement all the essential features of SNET presented in Section 1.

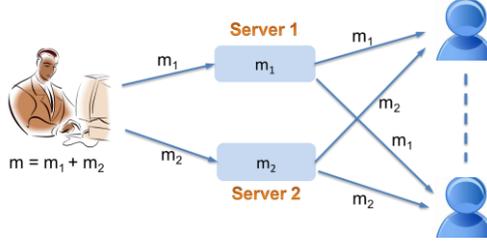

Figure 1: The Multi-Server Framework

A simple illustration of the system architecture is presented in Figure 1 where an SNET is managed by two servers or service providers. By utilizing two or more servers, symmetric encryption (such as AES) is no longer necessary to establish secure end-to-end communication. Instead, we can adopt a secret sharing scheme which can be at least as secure as AES depending on the underlying randomness. As shown in the figure, Bob's message $m$ is secretly represented with two shares, denoted by $m_1$ and $m_2$. The two shares are stored separately at Server 1 and Server 2. Since each share when viewed individually is random or pseudo-random (depending on the specific scheme), each alone does not leak any information regarding $m$. Then the servers can deliver their own shares to users connected to Bob in the network. By combining these shares locally, Bob's friends will be able to retrieve the actual message $m$. We will present three ways to generate these shares from simple to complex with increasing security guarantee in Sections 4.5 and 6.3. Here, we emphasize that this multi-server framework can achieve the following essential properties of an SNET:

- *Efficient and secure end-to-end communications*: As mentioned before, in a large group, AES based secure end-to-end communication places a large computational burden on the user. For example, suppose Bob wants to send a message to 100 of his friends. Then 100 unique, secure, end-to-end communications need to be established. On the other hand, in our system, Bob only needs to communicate with two servers securely. Under this particular example and based on a rough estimate, computations incurred on Bob's system could be 50 times more efficient under the proposed system architecture.

- *Oblivious communication*: Since the servers only need to perform very basic functionalities (e.g., managing user accounts, sending and receiving messages to and from their users), in theory, any service providers who allow their users to store and share data with other users can be used to implement the proposed SNET. In addition, if the SNET can be camouflaged into the existing services, it becomes hard for the service providers to detect the existence of the SNET. This allows our SNET to operate in an oblivious mode.

- *High degree of availability*: It is difficult for a third party to shut down the SNET service, because under the oblivious mode, even the servers do not know which users use their services as part of SNET. Also, if these servers provide daily essential services to the general public, it is difficult to shut down all these servers entirely in order to eliminate SNETs. Furthermore, in case a server is not functioning properly, dynamic configuration is possible to allow the users to choose another service provider to continue secure and oblivious communications with other users.

## 3.2 Technical Challenges

According to the system framework in Figure 1, although the general steps are straightforward, the actual implementation is not as simple as it seems. Before presenting implementation details, we first highlight the technical challenges. Our implementation will be centered around how to solve these challenges. We have two versions of implementation: Secret-Share and Secret-Share* secure under passive and active adversary models respectively. For Secret-Share and Secret-Share* to achieve all the security properties discussed in Section 3.3, there are several key technical challenges and components that need to be solved or developed:

1. *Generating secret shares for various data types:*
   Our goal is to support common data types, such as text, image, audio, video, etc. To simplify implementation, it is beneficial to have a unified way to generate secret shares for different data types.

2. *Optimization of data granularity*:
   Data granularity determines the computational efficiency. For example, to generate secret shares of a piece of text, we can use each character as a unit. On the other hand, two characters together could be considered as a single processing unit. Which one gives a better performance? This could depend on many factors, such as the types of pseudo-random number generators, big number libraries, and the underlying operating systems. Much work is required to discover the most efficient data granularity.

3. *Share synchronizations*:
   Since the cloud servers used by Secret-Share and Secret-Share* are independent, the way and the frequency that information is synchronized are different from one server to another. The Secret-Share application needs to combine corresponding secret shares together, so efficient ways have to be developed to link shares that are related to the same information. To solve this problem, we first need to investigate synchronization behaviors of each server. We also need to develop compact meta-data that enable correct linking among the secret shares for reconstructing the original information.

4. *Communication and storage minimization*:
   Not every user has an unlimited data plan. In addition, cloud storage providers only give free services up to a certain limit on storage capacity. Therefore, minimizing both communication and storage costs is crucial to attract users. To minimize communication cost, the Secret-Share application should only send delta changes of each file to the servers, but additional information is required to keep tracking these changes. The additional information could also increase the communication and storage costs. We will need to examine the best ways to save both costs at much as possible.

5. *Graphical user interface (GUI):*
   Easy to use GUI is essential, and the interface should allow users to send, receive, display and manage user data. The GUI should not be resource hungry.

6. *Preventing collusion*:
   As mentioned before, if the servers collude by sharing their secret shares, they will be able to discover the original information. If this happens, Secret-Share will no longer be privacy-preserving. In practice, since the servers are independent entities and their reputations worth billions of dollars. The chance for them to collude can be small. However, a TPA may force them to work together to discover the existence of an SNET and retrieve private user data. Efficient strategies are needed to prevent these malicious activities.

In what follows, Section 3.3 details our threat model and obliviousness, Section 4.2 addresses the first two technical challenges, Section 4.3 focuses on challenges 3 and 4, Section 4.4 discusses the GUI of Secret-Share, and methods to prevent collusions among the service providers and enforce security against active adversaries are proposed in Sections 4.5, 6.2 and 6.3.

## 3.3 The Threat Model and Levels of Obliviousness

To understand the security guarantee of SNET, we need to specify the parities involved and their adversarial behaviors. Since not all malicious behaviors can be prevented, it is important to clarify what can be prevented under SNET. Participating parties can be broken into three broad categories:

- SNET User: The goal of a user is to communicate with their family, friends, and others securely.

- Service provider (SP): An entity provides the basic data storage, user management, access control and inter-user communication functionalities.

- Third party authority (TPA): A third party that may interfere with the normal operations of IT industries.

Ideally, the security guarantee of an SNET needs to include the following properties:

- Obliviousness: The existence of SNET services should not be known to SP and TPA.

- Confidentiality: Message confidentiality should be preserved against SP, TPA and unauthorized users.

- Integrity: The messages should not be modified by unauthorized parties.

- Availability: The users should be able to use the SNET service whenever possible.

Next, we specify the adversarial behaviors of the participating parties, and with these behaviors, we emphasize which among the four properties listed above can be achieved by our system.

### 3.3.1 Adversary Models

When designing secure protocols, we generally consider two types of adversaries: passive and active. In the literature of SMC, they correspond to semi-honest and malicious adversaries. The passive adversary just follows the prescribed procedures of a secure protocol, but the adversary is allowed to use whatever he or she sees during the execution of the protocol is defeat the security guarantee. On the other hand, an active adversary can diverge arbitrarily from the normal execution of the protocol. When a protocol is secure against active adversary, it is also secure under the passive adversary model. We assume the following behaviors of the parties involved under the passive adversary model:

- TPAs: Just conduct their normal operations, and do not collude with SPs and users to discover the existence of SNET.

- SPs: Provide normal operations of their services. In addition, they do not collude with other entities. •

  SNET users: These users do not intentionally leak private communication data to unauthorized users.

The passive adversary assumption is consistent with a normal democratic society. It is reasonable to assume that no authority interferes with the freedom of speech and does not collude with SPs. In addition, reputable SPs do not have the incentive to collude to discover private user data to damage their priceless reputations. Under the active adversary model, we assume the following possible adversarial behaviors related to each party:
- TPAs: They may collude with SPs or SNET users to discover and identify the existence of SNET. They may also force SPs to stop their services.

- SPs: They may collude with other SPs, TPAs or SNET users to detect the existence of SNET, and to reconstruct the private communication messages among SNET users.

- SNET users: They may leak private communication data and the existence of SNET to other parties.

Clearly, to achieve obliviousness, confidentiality, integrity and availability is much easier under the passive adversary model. Regarding active adversaries, not every property can be guaranteed. As a result, we propose two different implementations, Secret-Share and Secret-Share*, of SNET under the passive and active adversary models respectively. In the following, we discuss what security properties can be achieved according to each model.

### 3.3.2 Security Guarantee in the Passive Adversary Model

Secret-Share is our implementation of SNET under the passive adversary model. Under this model, integrity and availability guarantees of Secret-Share directly depend on those provided by the SPs for their normal or regular users. Secret-Share can achieve confidentiality as long as the communication channels between the SPs and their users are secure. Secure channels between the SPs and their users do not imply secure end-to-end communication among the users which requires additional mechanisms. It can be tricky to attain obliviousness. For improved usability, the implementation of Secret-Share needs to use the application programming interface (API) of each SP. If the SP requires that whatever application uses its API, the application needs to disclose to the SP the kind of service the application provides. If this is the case, our current implementation of Secret-Share does not achieve obliviousness. On the other hand, if the SP does not require Secret-Share to disclose its service type, obliviousness can be attained. Regardless the SP's requirement of its API usage, Secret-Share* can guarantee obliviousness.

### 3.3.3 Security Guarantee in the Active Adversary Model

Secret-Share* is our implementation of SNET under the active adversary model. First, we show how obliviousness determines the other security properties. In this adversary model, the goal of the TPA is to discover the existence of SNET and to interrupt its service. The SPs may also want to discover the existence of SNET and to influence its operations. The users may disclose information received under an SNET to other entities, and they may also collude with TPA and SP to help them to discover hidden SNET services.

In order to achieve oblivious communication, the servers must be prevented from knowing that they are actually providing services for SNET. As a result, Secret-Share* must look indistinguishable from any other use of the service. To do this we do not use APIs provided by any SPs. The level of obliviousness achieved by Secret-Share* is summarized below:

- A TPA cannot easily identify the users of Secret-Share* and prevent these users from using the application. Also, the TPA does not know which SPs provide the back-end services for Secret-Share*.

- The SPs themselves do not know if their servers are utilized by the Secret-Share*, and cannot identify any of their users are part of the Secret-Share* user group.

- Obliviousness is impossible to attain if the users of Secret-Share* collude with either an TPA or SP.

The last point emphasizes a scenario equivalent to insider threat which cannot be prevented. However, if Secret-Share* is detected and shut down, users can still dynamically change its server settings to restart the oblivious communication services. Implementation details of both Secret-Share and Secret-Share* are presented in Section 4. Achieving absolute obliviousness is not possible. However, our implementations accomplish the best possible outcomes in today's computing environment. We will point out the potential limitations of our current implementations against more advanced but hypothetical attacks and possible strategies to eliminate them in Section 6.

Once obliviousness is achieved, the rest of the security properties: confidentiality, integrity and availability can also be achieved as long as the SPs do not intentionally stop serving their customers, modify user data and providing secure channels between the severs and customers. Assuming the SPs continue to provide credible services to their regular customers is reasonable since their incentive is to make money, and failure to follow through on service guarantees will quickly hurt their ability to do so.

# 4 Implementation Details of Secret-Share and Secret-Share*

Based on publicly available resources, one of the main goals of this paper is to show how to implement two instances or versions of SNET based on the system architecture given in Figure 1. The two versions are denoted by Secret-Share and Secret-Share*, which are secure in the passive and active adversary models respectively. The implementation of Secret-Share* is built on top of Secret-Share, and the key differences between the two version are highlighted in Table 2.

Table 2: Secret-Share vs. Secret-Share*

|  | Secret-Share | Secret-Share* |
| --- | --- | --- |
| SP choice | Fixed | Dynamic |
| Adversary | Passive | Active |
| Ease of use | High | Low |

## 4.1 Implementation Overview

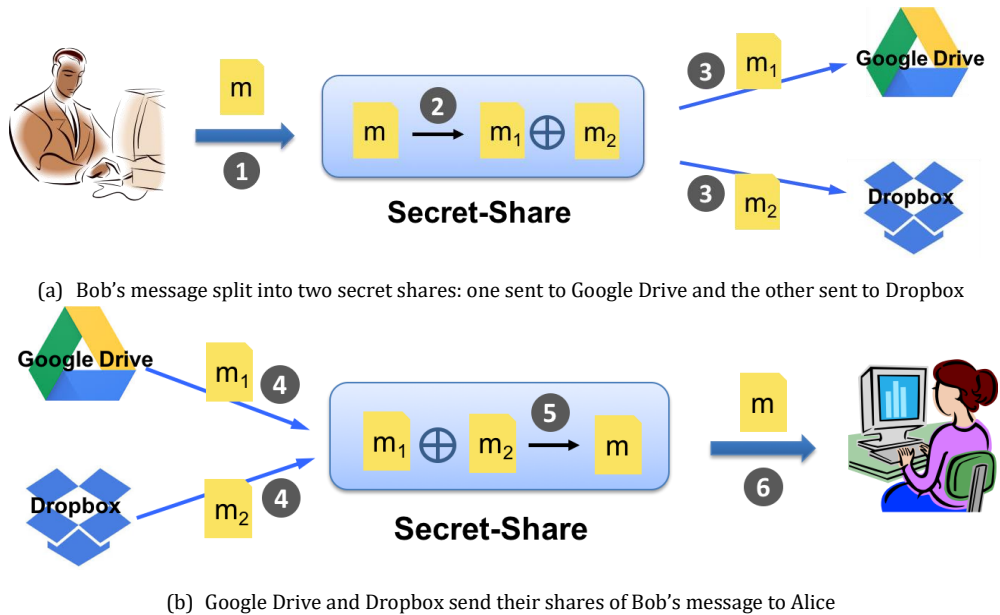

(a) Bob's message split into two secret shares: one sent to Google Drive and the other sent to Dropbox

(b) Google Drive and Dropbox send their shares of Bob's message to Alice

Figure 2: Overview of secure end-to-end communication between Alice and Bob

The main functionality of Secret-Share and Secret-Share* is to provide efficient and secure end-to-end user communication using free services from the cloud servers. The key differences between WhatsApp [34] and other social networking platforms [35, 27] are highlighted below:

- A user's profile and activities cannot be easily monitored by the service providers. This minimizes the adverse impact collusion and oversight a third-party authority can have.

- For Secret-Share*, the users have the flexibility to select a subset of cloud servers to store and manage their data. In addition to preventing specific cloud providers from having the full control of user data and activities, it also helps users take advantage of free cloud services.

- Secret-Share* application is oblivious to the cloud servers which further increases the difficulty for the service providers to monitor and track user activities.

- To implement Secret-Share and Secret-Share*, cloud servers only need to have data storage and sync capabilities among user local storages. Since these capabilities are part of free services for most clouds up to a certain storage limit, it is feasible for us to make Secret-Share freely available to the public without direct support from the clouds, and the cost incurred at the user side can also be minimal.

## 4.2 Share Generation and Distribution in Secret-Share

| $m$ | 01001100 | 01101001 | 01100110 | 01100101 | 00100000 | 01101001 | 01101001 | 00100000 | 01100111 | 01101111 | 01101111 | 01100100 |
|---|---|---|---|---|---|---|---|---|---|---|---|---|
| | L | i | f | e | Space | i | s | Space | g | o | o | d |
| $m_1$ | 10010101 | 01010011 | 01111101 | 11110000 | 10101011 | 10011100 | 10111100 | 10101100 | 10110101 | 10111110 | 11101001 | 10110111 |
| | • | S | } | ð | ≪ | oe | $\frac{1}{4}$ | ¬ | µ | $\frac{3}{4}$ | é | · |
| $m_2$ | 11011001 | 00111010 | 00011011 | 10010101 | 10001011 | 11110101 | 11010101 | 10001100 | 11010010 | 11010001 | 10000110 | 11010011 |
| | Ù | : | Escape | • | ⟨ | õ | Õ | OE | Ò | Ñ | † | Ó |

Table 3: Sample Text and its Secret Shares

Figure 2 shows the key components in the implementation of Secret-Share where Google Drive and Dropbox are used to replace the two servers given in Figure 1. In theory, any two cloud storage providers can be used by the users as long as two users share the same providers. However, in our current implementation, the two servers are fixed to Google Drive and Dropbox.

To send a secret message $m$ to Alice, first the Secret-Share application installed on Bob's device generates a pseudo-random bit sequence to represent the first secret share $m_1$ with the same size as $m$. Then the second share $m_2 = m_1 \oplus m$. As long as $m_1$ is random or pseudo-random, $m_2$ is also random or pseudo-random when viewed alone due to the property of the xor operator. In this illustration, $m_1$ is sent to Google Drive, and $m_2$ is sent to Dropbox. When both shares have permissions to be shared with Alice, $m_1$ and $m_2$ (or updated $m_1$ and $m_2$) will be delivered to Alice. Then, the Secret-Share application on Alice's device xors $m_1$ and $m_2$ to produce $m$. The GUI of Secret-Share will display $m$ to Alice.

To generate $m_1$, AES with stream cipher mode can be used. Alternatively, a secure pseudo-random number generator can also be used to generate $m_1$. The confidentiality guarantee of $m$ is directly based on the concept of one-time pad [29]. In our case, $m_1$ serves as the encryption key and $m_2$ is the ciphertext. Suppose that $s$ is the seed to generate $m_1$, then $s$ can be stored on Google Drive instead of $m_1$ to save storage space and bandwidth. Nonetheless, local computation on Alice's end is more expensive, and it becomes more difficult to synchronize the shares. Thus, the implementation of Secret-Share uses $m_1$ directly as the key to encrypt $m$.

### 4.2.1 How to Secretly Share any Information

Since digital data stored in any computers are just 0s and 1s, Secret-Share treats each message (e.g., text, image, video, etc.) as a sequence of bits. For instance, let $m$ denote a piece of text: *Life is good*. Under UTF-8 or ASCII code, $m$ can be represented as a binary string given in Table 3. As stated before, we can generate a pseudo-random sequence $m_1$ having the same size as $m$. Then $m_2$ is produced by $m_1 \oplus m$. This approach can be easily generalized to handle other data formats. To be secure against active adversaries, different secret sharing schemes are needed which will be discussed in Section 4.5.

## 4.3 Share Synchronization and Management

We have developed a preliminary version of Secret-Share desktop application. One of the key challenges is to efficiently manage and synchronize the secret shares. More specifically, we have to deal with how to (1) store a message and its secret shares, (2) minimize communication cost, (3) synchronize secret shares, and (4) prevent concurrent changes on synchronized messages.

In our current design, we store the corresponding messages for a group of users in a directory. The minimum group size is 2. In the directory, text-based messages for each direction are stored in a single file to reduce the number of data items and meta-data stored on the SP severs. For instance, Bob's messages to a group are kept in a single file. In addition, Bob keeps local copies of files for messages from other users to the group. Suppose the group has three users: Bob, Alice and Carl. Bob locally manages three text files for the group: one file stores the messages he sent to the group, and the other two files store the messages within the group from Alice and Carl respectively. For multimedia data (e.g., video, music, and pictures), each one is a single data item in the directory. They are managed similarly as the text files. For each directory $D$, we use to two separate directories $D_1$ and $D_2$ to store the secret shares of all data items in $D$ with the same file structures as $D$. In particular, the file names remain the same among the three directories.

### 4.3.1 Share Synchronizations

Following the previous notations, suppose $d$ is a file in $D$, and its secret shares $d_1$ and $d_2$ are stored in $D_1$ and $D_2$ respectively. Next, we briefly describe how $d$, $d_1$ and $d_2$ are synchronized with the SP servers under Secret-Share. Share synchronizations under Secret-Share* is very similar to Secret-Share except that $d_1$ and $d_2$ have local copies as well. We classify $d$ into either sending file or receiving file, and the owner of $d$ is the one who created the file. Considering user Bob, $d$ is a sending file if it stores messages from Bob to a group. On the other hand, $d$ is a receiving file if it stores messages from group members to Bob. How files are synced depends on the type of a file and state of its shares. The following steps highlight the file synchronization process in Secret-Share or Secret-Share*:

1. *d is a sending file*

    - When $d$ is created, $d_1$ and $d_2$ will be generated and stored on the SPs.
    - When $d$ is updated, the delta changes of $d_1$ and $d_2$ will be created and propagated to the SPs.
    - When $d$ is deleted, $d_1$ and $d_2$ will be deleted from the SP servers.

2. *d is a receiving file*

    - When $d_1$ and $d_2$ are created but $d$ does not exist, $d_1$ and $d_2$ are retrieved from the SP servers to generate a local copy of $d$.
    - When $d_1$ and $d_2$ are updated, the delta changes of $d_1$ and $d_2$ will be calculated to update $d$.
    - When $d_1$ and $d_2$ are no longer exist on the SP servers but $d$ still exists locally, $d$ will be deleted.

To keep tracking the file changes, we developed an index structure stored in the local device. Each entry of the index structure maps $d$ to $d_1$ and $d_2$ along with the last sync time among the three. By comparing the last sync time with the time stamps of $d_1$ and $d_2$ collected from the SP servers and the type of $d$, we are able to distinguish the six conditions listed before and perform the required actions to update these files. However, the actual implementations are quite complicated due to the limited functionalities provided by the APIs and not every file on the SPs are part of Secret-Share conversations.

## 4.4 Graphical User Interface

In the section, we provide several screen shots to show our initial implementation of the graphical user interface (GUI) for Secret-Share and Secret-Share*. We used the Qt [24] library for creating the GUIs, the GMP [11] library for generating pseudo-random sequences, and the Boost [6] library for manipulating files.

Before we can use the Secret-Share application, we need provide some configuration information as shown in Figure 3. We need a directory in which we can install the application, a directory to store unencrypted conversations locally, and a directory for each cloud that stores secrets so we can temporarily download their secret shares. Additionally, the application must be given permission to make file changes on a user's account. This is done via the Oauth2 authentication protocol so a user just has to log in when the browser pops up and provide permission as shown in Figure 4.

After the initial authentication with the cloud servers, Bob can start sharing conversations with Alice by clicking "new" button to establish a conversation (see Figure 5). This menu creates a local directory for the conversation, SampleConversation. This directory and its contents will be shared with "username" specified at each service providers. Alice will then need to add the shares through Secret-Share application installed on her device ( see Figure 6). The "add" button will poll each storage service for directories shared. The Secret-Share application downloads the messages in the selected directory and decrypts it. After that a conversation between Alice and Bob will be established, and their initial conversation is shown in Figures 7 and 8.

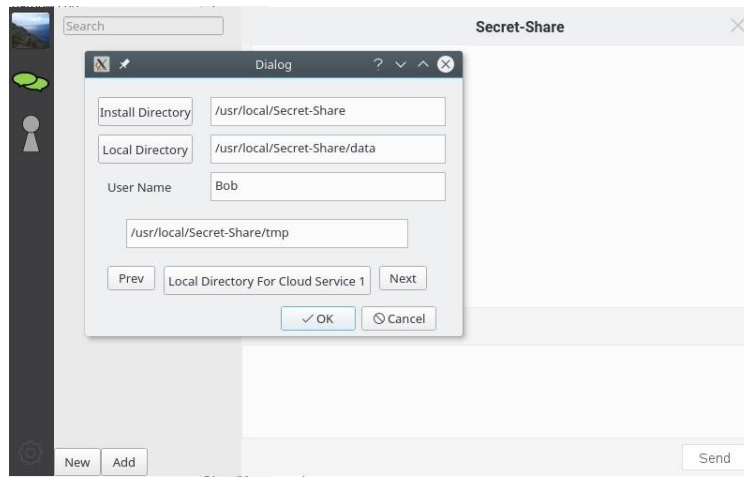

Figure 3: Configuration info

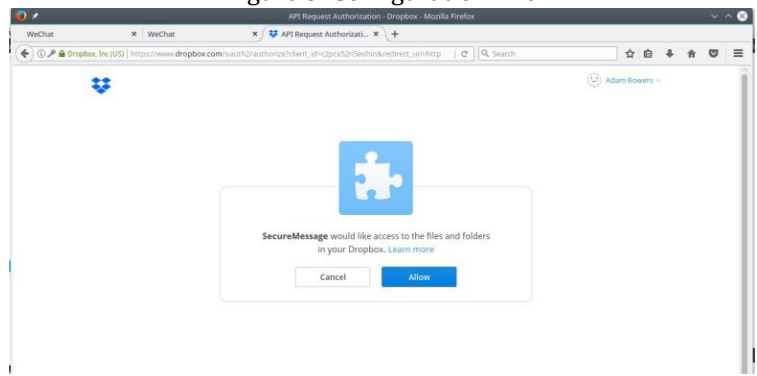

Figure 4: Oauth2 Authentication

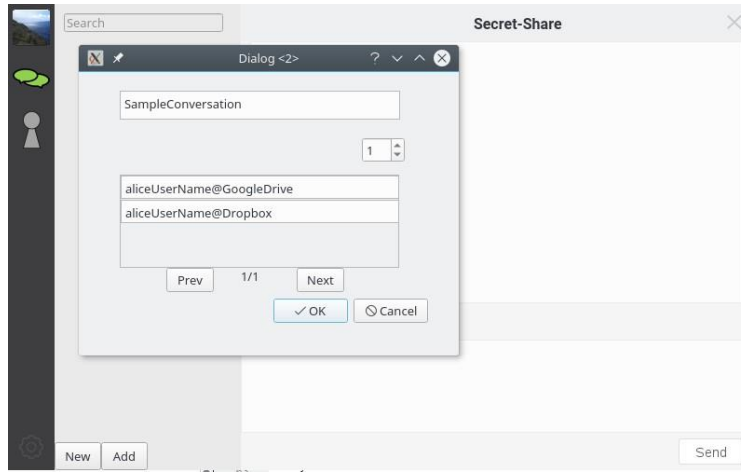

Figure 5: Share a conversation with a user

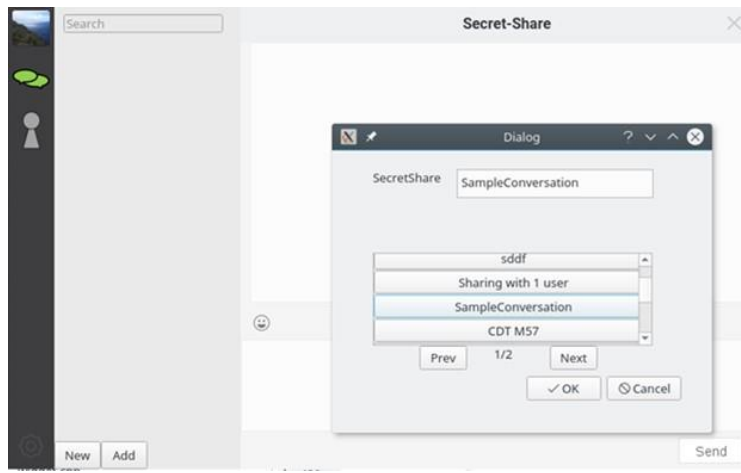

Figure 6: Add Shared conversation to the system

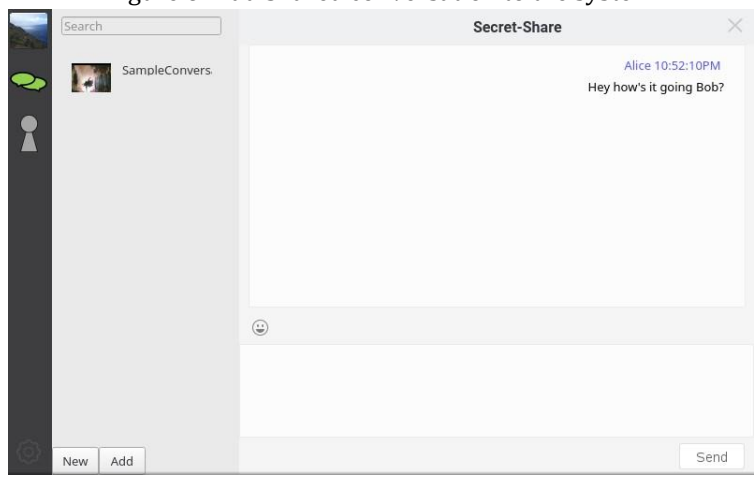

Figure 7: Alice Conversation Screen

## 4.5 Dynamic Server Selection for Secret-Share*

The implementation of Secret-Share is only secure against passive adversary mainly because the Secret-Share application may not be oblivious and collusion between the two servers can reveal user private data. The risk of being non-oblivious comes from the fact that the implementation of Secret-Share uses the APIs of Google Drive and Dropbox to improve usability. Therefore, the application ID of Secret-Share needs to be registered with both service providers. Based on this application ID, it is not difficult for the service providers to discover the actual services provided by Secret-Share. In addition, whenever a user starts the Secret-Share application, when authentication occurs with the two servers, either Google Drive or Dropbox will know actual users of Secret-Share. In this section, we present approaches to make Secret-Share oblivious and secure against active adversaries.

Here we highlight the key differences between Secret-Share and Secret-Share*. To provide a stronger security guarantee, the Secret-Share* application utilizes the following additional strategies:

- Avoid using the APIs of the service providers.

- Dynamic server selection among any two or more existing cloud service providers to achieve oblivious-

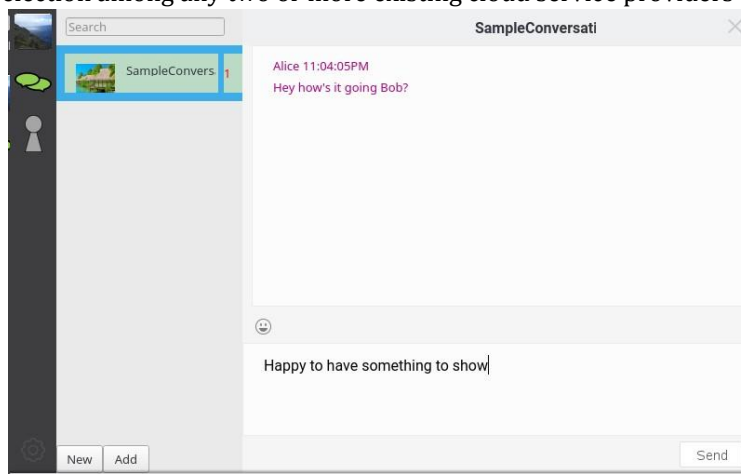

Figure 8: Bob Conversation Screen

ness.

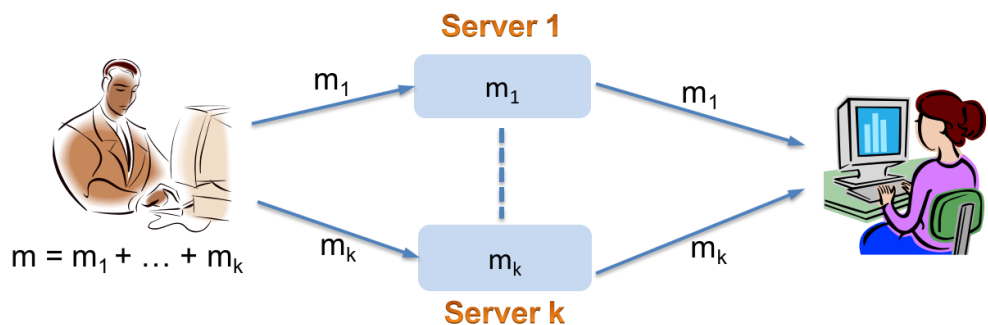

Figure 9: The *k*-Server Setting

In order to achieve obliviousness, first of all, we cannot use the APIs from the service providers to prevent them from tracking the use of the Secret-Share* application. Our current implementation of Secret-Share* still adopts Google Drive and Dropbox as the service providers, but the Secret-Share* application does not need to be authenticated with both servers for the users to utilize their services. Instead, the user device requires to install the Google Drive and Dropbox applications. The folders to store secret shares of the messages are kept in sync with the cloud storage by these local applications. The Secret-Share* application has the access to these folders to manage and update the secret shares locally.

### 4.5.1 The Number of Servers

As stated previously, the minimum number of servers required by our system architecture to implement efficient and secure end-to-end communication is two. However, to make collusion among the service providers more difficult, more than two servers can be used as long as they provide local applications to access their services. Suppose there are $k$ servers, we need to produce $k$ secret shares of the message $m$. First, we generate $k - 1$ pseudo-random sequences denoted by $m_1,...,m_{k-1}$. Then the share $m_k$ can be derived by $m_1 \oplus \cdots \oplus m_{k-1} \oplus m$. Each share will be stored and distributed by one of the servers. The message can be sent and received the same way as the two-server setting. In this way, all $k$ servers need to collude to discover $m$. Thus, the bigger the $k$, the harder for the servers to collude.

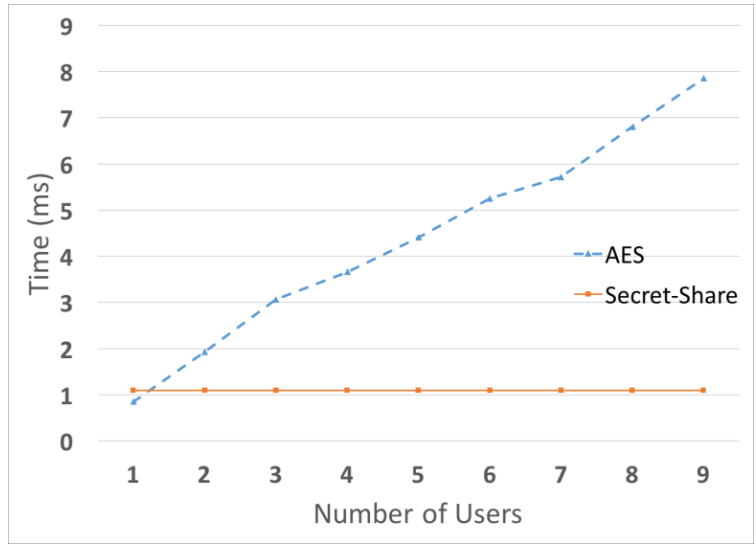

Figure 10: Computational efficiency Secret-Share vs. AES based implementation

Moreover, let us assume there are $n$ total service providers where $n \geq k$ and $k \geq 2$. The Secret-Share* application allows the user to configure the application with any $k$ of these $n$ service providers. Suppose Bob selects $k$ servers for his configuration of the Secret-Share* application, then the users who he communicates with also need to select the same $k$ servers. This introduces some randomness in the sever selection process and makes server collusion more difficult. As the number of servers increases, the storage, local computation and communication costs also increase, as well as degrading the ease of use. For example, in our current implementation, starting a new conversation has to be done manually. That is, Bob needs to individually invites Alice using each SP's invitation feature. After that, Bob and Alice can communicate freely under the Secret-Share* application.

# 5 Security Analysis and Performance Evaluation

The security guarantee of Secret-Share directly depends on how the shares of the original messages are generated. As long as one of the shares is random or pseudo-random and the xor operator is applied properly, Secret-Share will be secure against passive adversaries. More specifically, the confidentiality of the message is preserved by its secret shares and each SP only stores one of the shares. The integrity and availability can also be preserved as long as the SPs provide standard operations to their customers. The obliviousness may or may not be achieved. If the SP requires Secret-Share to fully disclose the services it provides when registering with the SP to use the SP's API, the obliviousness cannot be achieved. In addition, since each time Secret-Share is activated, it has to be authenticated with the SP due to the use of the APIs. As a consequence, the users can be tracked by the SPs.

To achieve obliviousness, our second implementation Secret-Share* eliminates the need of the APIs. Instead it only requires the local applications of the SPs to be installed on the user device. Secret-Share* also has the option of dynamically selecting a subset of SPs to support Secret-Share* functionalities. Combining these measures, Secret-Share* is able to achieve obliviousness, but it sacrifices some level of usability as mentioned before. Although we are not aware any existing attacks to break the obliviousness of Secret-Share*, we speculate on some potential strategies to detect the existence of Secret-Share* if TPAs and SPs were to work together. We will also propose counter measures to further secure Secret-Share* against the attacks. Detailed discussions on these issues are provided in Section 6.

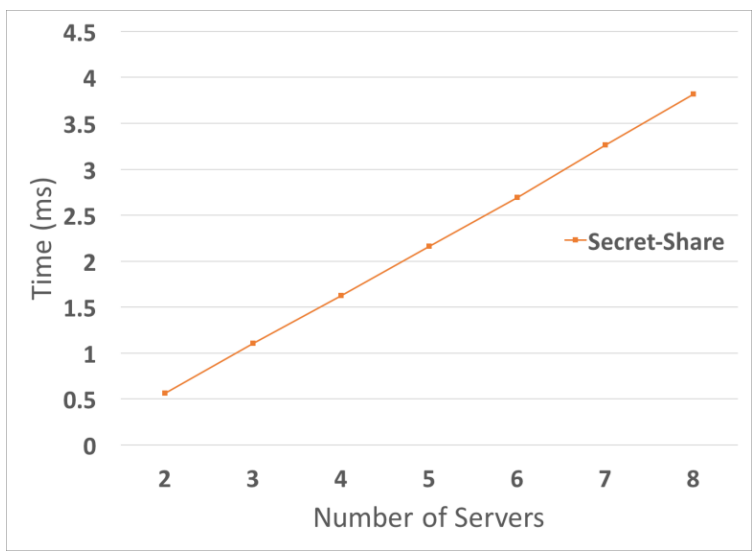

Figure 11: Computation cost with increasing number of servers

## 5.1 Efficiency of Secure End-to-End Communication under Secret-Share

As stated in Section 3.1, the general system architecture presented in Figure 1 leads to more efficient implementation of secure end-to-end communication within a group of users. Since we cannot access the source code for WhatsApp, we simulated the cost of the existing way to achieve secure end-to-end communication. Here we only focus on the local computation cost since the cost at the SPs is irrelevant to our implementations.

For both Secret-Share and Secret-Share*, the computation cost depends on the number of servers. Without loss generality, let treat the encryption operation as the basic computing unit (assuming pseudo-random number generation has similar complexity as AES)[1], *k* denotes the number of servers and *w* denote the number of users. To encrypt one message under Secret-Share, it requires $k-1$ encryptions and $k-1$ xor operations. Under the existing method for secure end-to-end communication, we need to perform *w* encryptions. For Secret-Share and Secret-Share*, the number of servers is constant; thus, it is apparent that our implementation provides a much more efficient way to achieve secure end-to-end communication when *w* is much larger than *k*. The results shown in Figure 10 are consistent with our theoretical analysis, where we adopted the standard AES functionality from the Crypto++ library [8] to implement the existing way to establish secure end-to-end communication, and *k* is fixed to 2 in the figure. The local computation cost of Secret-Share and Secret-Share* does not depend on the number of users in a group. However, as shown in Figure 11, when the number of servers increases, the local computation cost grows linearly.

## 5.2   Storage and Communication Costs

The storage requirement for Secret-Share and Secret-Share* can be higher than the existing solutions because each share needs its own storage. Again, let *k* denote the number of servers. Potentially, Secret-Share and Secret-Share* can require *k* times more storage than the existing solution. Since *k* is constant and equal to 2 in Secret-Share, this additional storage cost is negligible compared to the gain in local computation cost.

We estimate the communication cost be counting the number of messages sent and received given the original message *m* and a group of *w* users to share *m*. Under Secret-Share and for Bob to broadcast *m* in the group, *k* messages (each of which has the same size as *m*) are sent to the servers. Under the existing solution, Bob needs to send *w* messages. On the receiving end, the user needs to receive *k* messages in Secret-Share, but only one message when adopting existing solution. Because *k* is very small, the communication cost for Secret-Share is generally less than the existing solution by a factor of $\frac{w}{2k}$ for one send-receive cycle. The same result applies to Secret-Share*.

# 6   Limitations and Alternative Design

The implementation of Secret-Share* does not use the APIs of cloud servers, and server selection process can also be randomized among the existing service providers. As a result, Secret-Share* achieves a high degree of obliviousness. In this section, we discuss some advanced ways the service providers may still be able to discover the existence of Secret-Share*. Based on these attacks, we present the counter-measures to prevent these attacks or mitigate their negative impacts on the security guarantee of Secret-Share*.

The attacks discussed next assuming the worst case scenario where the current security measures adopted in Secret-Share* are no longer effective against certain active adversarial behaviors. To begin with, we point out the potential ways the existence of Secret-Share* can be discovered and what the main consequences are. To our knowledge, because Secret-Share* is the first secure communication tool to offer obliviousness, we are not aware of any existing methods to break its obliviousness. Thus, the following discussions on potential attacks are a little bit hypothetical.

## 6.1   Advanced Attacks against Obliviousness

When a user of Secret-Share*, say Bob, is malicious (insider threat), Bob can disclose messages he received to anyone. He can even disclose which cloud servers were utilized by his Secret-Share* application and the user IDs who have communicated with him. Nevertheless, Bob's behavior only affects the security guarantee of the users in communication with him through Secret-Share*, and the other Secret-Share* users will not be affected.

---

[1] In our actual implementation, we utilized the pseudo-random generator (PRGN) in the GMP library [11] to generate pseudo-random sequences, and we observed that using PRGN and xor to encrypt a fix message *m* is about three times faster than AES with CryptoPP [8].

When Bob is malicious, there is nothing we can do to prevent Bob from defeating the obliviousness of Secret-Share* for his group of users.

Like any insider threats, they cannot be prevented. Thus, we assume Secret-Share* users do not collude with service providers (SPs) and third-party authorities (TPAs). Under this assumption, we emphasize what SPs and TPAs could do to discover the existence of Secret-Share* by hypothesizing the following methods:

- *Discovery of random messages*: Since the secret shares of $m$ are pseudo-random bit sequences, it may be possible to detect these messages via outlier detection algorithms. An attack algorithm could build a general user profile to model message characteristics of a regular user of a cloud storage. This model may be able to predict or detect pseudo-random messages. After detecting a specific user, say Bob, sharing a large amount of pseudo-random messages, the SP can further verify messages of the users in communication with Bob. These users may lead the SP to other group of users to infer the existence of Secret-Share*. Even if this attack can be successfully carried out, a single SP still cannot discover the original messages shared among the identified users. However, a group of SPs may be able to collaboratively discover a common set of users to reconstruct their original messages.

- *Network traffic analysis*: Another possible way to detect the existence of Secret-Share* is to examine the network traffic. Suppose a TPA has access to the network managed by Bob's Internet service provider (ISP). If the TPA observes that Bob simultaneously accesses multiple SPs, Bob could be a user of Secret-Share*. By work with the SPs, the TPA can verify its suspicion about Bob.

## 6.2 The Counter Measures

The attacks discussed previously are possible in theory, but can be extremely difficult to carry out in practice. On the other hand, even in the worst case, we have counter measures to eliminate or mitigate these attacks.

- *Randomization and steganography*: To prevent data mining and machine learning algorithms to detect random messages, we can "randomize" the random messages by embedding them into normal messages or files. This can distort the patterns the attack algorithm focuses on. The embedding algorithms need to be built into the Secret-Share* application. In addition, the embedding process must be invertible to retrieval the embedded messages. Another counter measure is to adopt steganography techniques [3], where the secret shares can be hidden in an image or other multimedia data. Why not embed the original message $m$ directly? The main reason is that when steganography techniques fail, we still have an additional layer of protection.

- *Prevention of traffic analysis*: One of the best tools to combat network traffic analysis is Tor [31], which can hide the destinations of network packages from the ISPs. Tor needs not to be built into Secret-Share* as long as Bob's device supports the Tor software. Before using Secret-Share*, Tor needs to be activated first.

Network security is like "cat and mouse" game, and the above counter measures may fail to deter the attacks. Therefore, we discuss additional approaches in case the counter measures are no longer effective. That is, by colluding, the SPs may reconstruct the original messages of Secret-Share* users.

## 6.3 Advanced Secret Sharing Schemes

As discussed in Section 4.5, by increasing the number of servers, the collusion among the SPs becomes more and more difficult. However, it may not be possible to find more than two SPs. If this is the case, we can adopt more advanced secret sharing schemes to prevent collusion: additive or threshold-based schemes.

Assuming $s$ is a non-negative integer, to secretly share $s$, first randomly select a value $s_1$ from $Z_N$. Then $s_2$ can be computed according to $s = s_1 + s_2$ mod $N$. Like before, the secret shares $s_1$ and $s_2$ can be stored at two non-colluding and independent servers. If Alice has the permissions to access to the shares, the two servers will send $s_1$ and $s_2$ to Alice. Combining the two secrets together, Alice can learn $s$. Since $s_1$ is randomly generated, it does not leak any information regarding $s$. In addition, because of the + and mod operations, $s_2$ is also randomly distributed in $Z_N$. That is, $s_2$ alone does not reveal any information regarding $s$. Thus, in cryptography, $s_1$ and $s_2$ are called secret shares of $s$.

To generate shares for any message types, first we need to represent each message in its binary format as shown in Table 3. Once $N$ is fixed, we can determine the appropriate block size to generate the shares. For example, if $N = 2^8$, then 8-bit or one byte is the block size. The message $m$ is divided into $\frac{|m|}{|N|}$ blocks, and each block can be secretly shared using the scheme described above. This scheme is very efficient and only requires storage capability of the servers. The application running on the client side is very lightweight. More importantly, the end-to-end communication or information sharing is as at least secure as AES based implementations [13] depending on the underlying randomness.

In the $k$-server setting, similar to the example given in Section 4.5.1, we can easily extend the two-share computations to $k$ shares. First, randomly select $k - 1$ shares, and then derive the last share by the relationship of $s = s_1 + \cdots + s_k$ mod $N$. When $N$ is known to the cloud servers, the security guarantee of this scheme is the same as the xor based approach against colluding servers. However, when $N$ is selected randomly within a certain range, even if the servers collude, they still cannot properly reconstruct the message. For this scheme to work, the users who communicate among themselves need to agree on the same $N$ beforehand. Additionally, we can adopt Shamir's secret sharing scheme.

### 6.3.1 Shamir's Secret Sharing

A well known threshold based secret sharing scheme was proposed by Shamir [28]: an algorithm for dividing a secret into parts among multiple participants such that each participant gets its own unique part. The algorithm requires either some of the parts or all of them to reconstruct the secret. The scheme allows to set a threshold on the number of parts required to construct the secret. To create a ($t,n$) threshold scheme to share a secret $S \in F$, where $F = Z_p$ for some prime $p$, we can follow the steps given below:

- Choose random $t - 1$ coefficients $r_1,...,r_{t-1} \in F$
- Construct polynomial, $f(x) = r_0 + r_1x + r_2x^2 + \cdots + r_{t-1}x^{t-1}$, where $r_0 = S$
- Compute shares $s_i = (i, f(i))$, for $i = 1,...,n$.

Given at least $t$ of these shares indexed by $\{i_1, i_2,...,i_t\}$, the secret can be re-constructed. The parameter $p$ serves a similar purpose as $N$ in the additive secret sharing scheme, and the way to apply Shamir's secret sharing in Secret-Share* is the same as that of additive secret sharing.

Although the Shamir scheme is computationally more expensive, it does have an additional advantage by achieving fault-tolerance. Suppose there are $n$ SPs. Under the additive secret sharing, the failure of one SP will prevent Secret-Share* from sending and receiving new messages. On the other hand, under Shamir, as long as $t$ out of $n$ SPs function properly, Secret-Share* will not be affected.

## 6.4 A Note on Man-in-the-Middle Attack

Our current system does not consider man-in-the-middle as a potential threat as long as we communicate with established services such as Google Drive and Dropbox. This is because those services already offer ways (based on the public key infrastructure) to communicate securely and not susceptible to the main-in-the-middle attack. The local applications that sync files handle communicating with the services in the same secure way. If our application handles syncing files, we still communicate with secure endpoints provided by the service.

Network analysis of our uploads might yield information, but Tor or [16] could be integrated into our application to prevent network traffic analysis.

# 7 Conclusion and Future Work

In this paper, we first discussed the need of a stealth (i.e., secure and oblivious) communication network SNET that provides efficient and secure end-to-end communication, and whose existence cannot be easily detected by the service providers used or third-party authorities. As a result, SNET provide a more secure communication platform for users to exchange messages without disclosing sensitive contents to the service providers. Being oblivious, a third party may not be able to shut down SNET services even collaborating with the service providers. To show the feasibility of SNET, we presented the implementation details of two instances of SNET, namely Secret-Share and Secret-Share*. The Secret-Share application is secure against passive adversaries; whereas, Secret-Share* is secure against active adversaries as well. To further improve the security of Secret-Share*, we proposed several strategies to prevent advanced attacks and collusions among service providers.

We have produced desktop applications for both Secret-Share and Secret-Share*. As part of our future work, we will develop mobile applications for Secret-Share and Secret-Share*. We also plan to improve the usability of Secret-Share* by automating the user invitation feature.